\documentclass[twoside,reqno]{lt7-proc}
\usepackage{epsfig,cite}
\usepackage{amssymb,amsmath}
\usepackage{times}
\newcommand\rf[1]{(\ref{eq:#1})}
\newcommand\lab[1]{\label{eq:#1}}
\newcommand\nonu{\nonumber}
\newcommand\br{\begin{eqnarray}}
\newcommand\er{\end{eqnarray}}
\newcommand\be{\begin{equation}}
\newcommand\ee{\end{equation}}

\newcommand\lb{\lbrack}
\newcommand\rb{\rbrack}

\newcommand\lcurl{\left\{}
\newcommand\rcurl{\right\}}
\renewcommand\({\left(}
\renewcommand\){\right)}
\newcommand\bv{\bigm\vert}               

\newcommand\bc{\begin{center}}
\newcommand\ec{\end{center}}

\newcommand\partder[2]{\frac{{\partial {#1}}}{{\partial {#2}}}}

\renewcommand\a{\alpha}
\renewcommand\b{\beta}

\newcommand\vareps{\varepsilon}
\newcommand\g{\gamma}
\newcommand\G{\Gamma}

\newcommand\h{\frac{1}{2}}
\renewcommand\k{\kappa}
\renewcommand\l{\lambda}

\newcommand\m{\mu}
\newcommand\n{\nu}

\newcommand\om{\omega}

\newcommand\vp{\varphi}
\renewcommand\P{\Phi}
\newcommand\pa{\partial}

\newcommand\pr{\prime}

\newcommand\s{\sigma}

\renewcommand\t{\tau}
\renewcommand\th{\theta}

\newcommand\cA{{\mathcal A}}

\newcommand\cF{{\mathcal F}}

\newcommand{\ct}[1]{\cite{#1}}
\newcommand{\bib}[1]{\bibitem{#1}}
\newcommand\PRL[3]{\textsl{Phys. Rev. Lett.} \textbf{#1}, #3 (#2)}
\newcommand\NPB[3]{\textsl{Nucl. Phys.} \textbf{B#1}, #3 (#2)}

\newcommand\PRD[3]{\textsl{Phys. Rev.} \textbf{D#1}, #3 (#2)}

\newcommand\PLB[3]{\textsl{Phys. Lett.} \textbf{#1B}, #3 (#2)}
\newcommand\CQG[3]{\textsl{Class. Quantum Grav.} \textbf{#1}, #3 (#2)}

\newcommand\IJMPA[3]{\textsl{Int. J. Mod. Phys.} \textbf{A#1}, #3 (#2)}

\newcommand\ydot{\stackrel{.}{y}}
\newcommand\yddot{\stackrel{..}{y}}

\begin{document}
\sloppy \raggedbottom
\setcounter{page}{1}

\newpage
\setcounter{figure}{0}
\setcounter{equation}{0}
\setcounter{footnote}{0}
\setcounter{table}{0}
\setcounter{section}{0}



\title{Lightlike Braneworlds}

\runningheads{Guendelman, Kaganovich, Nissimov, Pacheva}{Lightlike Braneworlds}

\begin{start}


\coauthor{Eduardo Guendelman}{1},
\coauthor{Alexander Kaganovich}{1},
\author{Emil Nissimov}{2},
\coauthor{Svetlana Pacheva}{2}

\address{Department of Physics, Ben-Gurion University of the Negev, P.O.Box 653, IL-84105
~Beer-Sheva, Israel}{1}
\address{Institute for Nuclear Research and Nuclear Energy, Bulgarian Academy of Sciences, 
Boul. Tsarigradsko Chausee 72, BG-1784 ~Sofia, Bulgaria}{2}


\begin{Abstract}
We propose a new class of $p$-brane models describing intrinsically {\em lightlike} 
branes in any world-volume dimensions. Properties of the dynamics of these
lightlike $p$-branes in various gravitational backgrounds of interest in
the context of braneworlds are briefly described.
Codimenion two (and more) lightlike braneworlds perform in their ground
states non-trivial motions in the extra dimensions in sharp contrast to
standard (Nambu-Goto) braneworlds.
\end{Abstract}
\end{start}


\section{Introduction}

Lightlike branes (\textsl{LL-branes}, for short) are of particular interest in 
general relativity primarily due to their role: (i) in describing impulsive lightlike 
signals arising in cataclysmic astrophysical events \ct{barrabes-hogan};
(ii) as basic ingredients in the so called ``membrane paradigm'' 
theory \ct{membrane-paradigm} of black hole physics; (iii) in the
context of the thin-wall description of domain walls coupled to 
gravity \ct{Israel-66,Barrabes-Israel-Hooft}.

More recently, \textsl{LL-branes} became significant also in the context of
modern non-perturbative string theory, in particular, as the so called
$H$-branes describing quantum horizons (black hole and cosmological)
\ct{kogan-01}, as well appearing as Penrose limits of baryonic $D(=$Dirichlet)
branes \ct{mateos-02}.

In the original papers \ct{Israel-66,Barrabes-Israel-Hooft} \textsl{LL-branes}
in the context of gravity and cosmology have been extensively studied from a 
phenomenological point of view, \textsl{i.e.}, by introducing them without specifying
the Lagrangian dynamics from which they may originate\footnote{In a recent paper 
\ct{barrabes-israel-05} brane actions in terms of their pertinent extrinsic geometry
have been proposed which generically describe non-lightlike branes, whereas the 
lightlike branes are treated as a limiting case.}. 
On the other hand, we have proposed in a series of recent papers 
\ct{will-brane-all} a new class of concise Lagrangian
actions, among them -- {\em Weyl-conformally invariant} ones, providing a
derivation from first principles of the \textsl{LL-brane} dynamics.
The latter \textsl{LL-brane} actions were, however, limited to $(p+1)=${\em odd}
world-volume dimensions. 

In Section 2 of the present paper we extend our previous
construction to the case of \textsl{LL-brane} actions for {\em arbitrary}
world-volume dimensions. In Section 4 we discuss the properties of 
\textsl{LL-brane} 
dynamics in generic static gravitational backgrounds, in particular, the
case with two extra dimensions from the point of view of ``braneworld'' scenarios
\ct{brane-world} (for a review, see \ct{brane-world-rev}). 
Unlike conventional braneworlds, where the underlying branes are of Nambu-Goto type 
(\textsl{i.e.}, describing massive brane modes) and in their ground state they 
position themselves at some
fixed point in the extra dimensions of the bulk space-time, our lightlike
braneworlds perform in the ground state non-trivial motions in the extra
dimensions -- planar circular, spiral winding, \textsl{etc.} depending on the
topology of the extra dimensions. Finally, in the outlook section we briefly
outline the treatment of the special case of codimension one lightlike branes 
which play an important role in the context of black hole physics. Also we
comment on the role of lightlike branes in Kaluza-Klein scenarios with
singular bulk metrics \ct{searight}.

\section{Generalized Gauge Field Description of Lightlike Branes}

The main ingredients of our construction of \textsl{LL-brane} actions for
arbitrary $(p\!+\!1)$ world-volume dimensions are:

\begin{itemize}
\item
Alternative non-Riemannian integration measure density $\P (\vp)$ (volume form) on
the $p$-brane world-volume manifold:
\be
\P (\vp) \equiv \frac{1}{(p+1)!} \vareps_{I_1\ldots I_{p+1}}
\vareps^{a_1\ldots a_{p+1}} \pa_{a_1} \vp^{I_1}\ldots \pa_{a_{p+1}} \vp^{I_{p+1}}
\lab{mod-measure-p}
\ee
instead of the usual $\sqrt{-\g}$. Here $\lcurl \vp^I \rcurl_{I=1}^{p+1}$ are
auxiliary world-volume scalar fields; $\g_{ab}$ ($a,b=0,1,{\ldots},p$)
denotes the intrinsic Riemannian metric on the world-volume, and
$\g = \det\Vert\g_{ab}\Vert$.
\item
Auxiliary $(p-1)$-rank antisymmetric tensor gauge field $A_{a_1\ldots a_{p-1}}$
on the world-volume with $p$-rank field-strength and its dual:
\be
F_{a_1 \ldots a_{p}} = p \pa_{[a_1} A_{a_2\ldots a_{p}]} \quad ,\quad
F^{\ast a} = \frac{1}{p!} \frac{\vareps^{a a_1\ldots a_p}}{\sqrt{-\g}}
F_{a_1 \ldots a_{p}}  \; .
\lab{p-rank}
\ee
\end{itemize}

Note the simple identity:
\be
F_{a_1 \ldots a_{p-1}b}F^{\ast b} = 0 \; ,
\lab{F-id}
\ee
which will play a crucial role in what follows, and let us also introduce
the short-hand notation:
\be
F^2 \equiv F_{a_1 \ldots a_{p}} F_{b_1 \ldots b_{p}} 
\g^{a_1 b_1} \ldots \g^{a_p b_p} \; .
\lab{F2-id}
\ee

We now propose the following reparametrization invariant action describing
intrinsically lightlike $p$-branes for any world-volume dimension $(p+1)$:
\be
S = - \int d^{p+1}\s \,\P (\vp)
\Bigl\lb \h \g^{ab} \pa_a X^{\m} \pa_b X^{\n} G_{\m\n}(X) - L\!\( F^2\)\Bigr\rb
\lab{LL-brane}
\ee
using the objects \rf{mod-measure-p} and \rf{p-rank}, where $L\!\( F^2\)$ is
{\em arbitrary} function of $F^2$ \rf{F2-id} and $G_{\m\n}(X)$ denotes the
Riemannian metric of the bulk space-time. 

\textbf{Remark.} For the special choice
$L\!\( F^2\)= \( F^2\)^{1/p}$ the action \rf{LL-brane} becomes 
manifestly invariant under {\em Weyl (conformal) symmetry}: 
$\g_{ab}\!\! \longrightarrow\!\! \g^{\pr}_{ab} = \rho\,\g_{ab}$,
$\vp^{i} \longrightarrow \vp^{\pr\, i} = \vp^{\pr\, i} (\vp)$ with Jacobian 
$\det \Bigl\Vert \frac{\pa\vp^{\pr\, i}}{\pa\vp^j} \Bigr\Vert = \rho$.

Rewriting the action \rf{LL-brane} in the following equivalent form:
\be
S = - \int d^{p+1}\!\s \,\chi \sqrt{-\g}
\Bigl\lb \h \g^{ab} \pa_a X^{\m} \pa_b X^{\n} G_{\m\n}(X) - L\!\( F^2\)\Bigr\rb
\quad, \quad
\chi \equiv \frac{\P (\vp)}{\sqrt{-\g}}
\lab{LL-brane-chi}
\ee
we see that the composite field $\chi$ plays the role of a dynamical
(variable) brane tension.

The equations of motion obtained from \rf{LL-brane} w.r.t. measure-building 
auxiliary scalars $\vp^I$ and $\g^{ab}$ read, respectively:
\br
\h \g^{cd}\(\pa_c X \pa_d X\) - L\!\( F^2\) = M \;
\Bigl( = \mathrm{integration ~const}\Bigr) \; ,
\lab{phi-eqs} \\
\h\(\pa_a X \pa_b X\) - p L^\pr\!\( F^2\) F_{a a_1\ldots a_{p-1}}
\g^{a_1 b_1}\ldots \g^{a_{p-1} b_{p-1}} F_{b b_1 \ldots b_{p-1}} = 0 \; ,
\lab{gamma-eqs}
\er
where we have introduced short-hand notation for the induced metric:
\be
\(\pa_a X \pa_b X\) \equiv \pa_a X^\m \pa_b X^\n G_{\m\n}  \; .
\lab{metric-short-hand}
\ee
Let us note that Eqs.\rf{gamma-eqs} can be viewed as $p$-brane analogues of the
string Virasoro constraints.

Eqs.\rf{phi-eqs}--\rf{gamma-eqs} have the following profound consequences.
First, taking the trace in \rf{gamma-eqs} and comparing with \rf{phi-eqs} 
implies the following crucial relation for the Lagrangian function $L\( F^2\)$: 
\be
L\!\( F^2\) - p F^2 L^\pr\!\( F^2\) + M = 0 \; ,
\lab{L-eq}
\ee
which determines $F^2$ on-shell as certain function of the integration
constant $M$, \textsl{i.e.}
\be
F^2 = F^2 (M) = \mathrm{const} \; .
\lab{F2-const}
\ee

The second and most important implication of Eqs.\rf{gamma-eqs} is due to
the identity \rf{F-id} which implies that the induced metric 
\rf{metric-short-hand} on the world-volume of the $p$-brane model \rf{LL-brane} 
is {\em singular} (as opposed to the ordinary Nambu-Goto brane):
\br
\(\pa_a X \pa_b X\) F^{\ast b} = 0 \quad ,\quad \mathrm{i.e.}\;\;
\(\pa_F X \pa_F X\) = 0 \;\; ,\;\; \(\pa_{\perp} X \pa_F X\) = 0 \; ,
\lab{LL-constraints}
\er
where $\pa_F \equiv F^{\ast a} \pa_a$ and $\pa_{\perp}$ are derivatives 
along the tangent vectors in the complement of $F^{\ast a}$.

Thus, we arrive at the following important conclusion: every point on the 
surface of the $p$-brane \rf{LL-brane} moves with the speed of light
in a time-evolution along the vector-field $F^{\ast a}$. Therefore, we will
name \rf{LL-brane} by the acronym {\em LL-brane} (Lightlike-brane) model.

Before proceeding let us point out that we can add to the \textsl{LL-brane}
action \rf{LL-brane} natural couplings to bulk Maxwell $\cA_\m$ and Kalb-Ramond
$\cA_{\m_1 \ldots \m_{p+1}}$ gauge fields:
\br
S = - \int d^{p+1}\s \,\P (\vp)
\Bigl\lb \h \g^{ab} \pa_a X^{\m} \pa_b X^{\n} G_{\m\n} (X) - L\!\( F^2\)\Bigr\rb
\nonu \\
-q \int d^{p+1}\s \vareps^{ab_1\ldots b_p} F_{b_1\ldots b_p} \pa_a X^\m \cA_\m (X)
\nonu \\
-\frac{\b}{(p+1)!} \int d^{p+1}\s \vareps^{a_1\ldots a_{p+1}}
\pa_{a_1} X^{\m_1} \ldots \pa_{a_{p+1}} X^{\m_{p+1}}\cA_{\m_1 \ldots \m_{p+1}} \;.
\lab{LL-brane-ext}
\er
The additional coupling terms to the bulk fields do not affect
Eqs.\rf{phi-eqs} and \rf{gamma-eqs}, so that the conclusions about on-shell
constancy of $F^2$ \rf{F2-const} and the lightlike nature \rf{LL-constraints} of the 
$p$-branes under consideration remain unchanged.
The second Chern-Simmons-like term in \rf{LL-brane-ext} is a special case 
of a class of Chern-Simmons-like couplings of extended objects to external
electromagnetic fields proposed in ref.\ct{Aaron-Eduardo}.

The Kalb-Ramond gauge field has special significance in $D=p+2$-dimensional
bulk space-time. The single independent component $\cF$ of its field-strength:
\br
\cF_{\m_1\ldots\m_D} = D\pa_{[\m_1}\cA_{\m_2\ldots\m_D]} =
\cF \sqrt{-G} \vareps_{\m_1\ldots\m_D}
\lab{F-D}
\er
when coupled to gravity produces a dynamical (positive) cosmological constant
(cf. ref.\ct{Aurilia-Townsend} for $D\!=\!4$; recall, here $D\!=\!p+2$):
\be
K = \frac{8\pi G_N}{p(p+1)} \cF^2 \; .
\lab{dynamical-cc}
\ee

It remains to write down the equations of motion w.r.t. auxiliary
world-volume gauge field $A_{a_1 \ldots a_{p-1}}$ and $X^\m$ produced by the
action \rf{LL-brane-ext}:
\be
\pa_{[a}\( F^{\ast c} \g_{b]c}\, \chi L^\pr(F^2)\)
+ \frac{q}{4}\pa_a X^\m \pa_b X^\n \cF_{\m\n}(X) = 0
\; ;
\lab{A-eqs}
\ee
\br
\pa_a \(\chi \sqrt{-\g} \g^{ab} \pa_b X^\m\) + 
\chi \sqrt{-\g} \g^{ab} \pa_a X^\n \pa_b X^\l \G^\m_{\n\l}(X)
\nonu \\
-q \vareps^{ab_1\ldots b_p} F_{b_1\ldots b_p} \pa_a X^\n \cF_{\l\n}(X)G^{\l\m}(X)
\nonu \\
- \frac{\b}{(p+1)!} \vareps^{a_1\ldots a_{p+1}} \pa_{a_1} X^{\m_1} \ldots
\pa_{a_{p+1}} X^{\m_{p+1}} \cF_{\l \m_1 \dots \m_{p+1}}(X) G^{\l\m}(X) = 0 \; .
\lab{X-eqs}
\er
Here $\chi$ is the dynamical brane tension as in \rf{LL-brane-chi},
\br
\cF_{\m\n} = \pa_\m \cA_{\n} - \pa_\n \cA_{\m} \quad ,\quad
\cF_{\l \m_1 \dots \m_{p+1}} = (p+2) \pa_{[\l} \cA_{\m_1 \dots \m_{p+1}]}
\lab{F-strength}
\er
are the corresponding gauge field-strengths,
\be
\G^\m_{\n\l}=\h G^{\m\k}\(\pa_\n G_{\k\l}+\pa_\l G_{\k\n}-\pa_\k G_{\n\l}\)
\lab{affine-conn}
\ee
is the Christoffel connection for the external metric,
and $L^\pr(F^2)$ denotes derivative of $L(F^2)$ w.r.t. the argument $F^2$.

\section{Gauge-Fixed Equations of Motion}

Invariance under world-volume reparametrizations allows to introduce the
standard synchronous gauge-fixing conditions:
\be
\g^{0i} = 0 \;\; (i=1,\ldots,p)) \quad ,\quad \g^{00} = -1 \; .
\lab{gauge-fix}
\ee
In what follows we will also use a natural ansatz for the auxiliary world-volume
gauge field-strength:
\be
F^{\ast i}= 0 \;\; (i=1,{\ldots},p) \quad ,\quad \mathrm{i.e.} \;\;
F_{0 i_1 \ldots i_{p-1}} = 0 \; ,
\lab{F-ansatz}
\ee
the only non-zero component of the dual strength being:
\br
F^{\ast 0}= \frac{1}{p!} \frac{\vareps^{i_1 \ldots i_p}}{\sqrt{\g^{(p)}}}
F_{i_1 \ldots i_p} \; ,
\lab{F-ansatz-1} \\
\g^{(p)} \equiv \det\Vert\g_{ij}\Vert\;\; (i,j=1,\ldots,p) \quad ,\quad
F^2 = p! \( F^{\ast 0}\)^2 = \mathrm{const}  \; .
\nonu
\er
According to \rf{LL-constraints} the meaning of the ansatz \rf{F-ansatz} is
that the lightlike direction $F^{\ast a} \pa_a \simeq \pa_0 \equiv \pa_\t$,
\textsl{i.e.}, it coincides with the brane proper-time direction.
Biancchi identity $\pa_a F^{\ast a}=0$ together with 
\rf{F-ansatz}--\rf{F-ansatz-1} implies:
\be
\pa_0 F_{i_1 \ldots i_p} = 0 \;\; \longrightarrow \;\; 
\pa_0 \sqrt{\g^{(p)}} = 0  \; .
\lab{gamma-p-const}
\ee

Using \rf{gauge-fix} and \rf{F-ansatz} the equations of motion \rf{gamma-eqs},
\rf{A-eqs} and \rf{X-eqs} acquire the form, respectively:
\be
\(\pa_0 X\,\pa_0 X\) = 0 \quad ,\quad \(\pa_0 X\,\pa_i X\) = 0 \quad ,\quad
\(\pa_i X\,\pa_j X\) - 2a_1(M)\, \g_{ij} = 0
\lab{gamma-eqs-0}
\ee
(Virasoro-like constraints), where the constant:
\be
a_1(M) \equiv F^2 L^\pr (F^2)\bv_{F^2 = F^2(M)}
\lab{a1-const}
\ee
(it must be strictly positive);
\be
\pa_i \chi + \frac{q}{a_2(M)}\pa_0 X^\m \pa_i X^\n \cF_{\m\n} = 0 \quad ,\quad
\pa_i X^\m \pa_j X^\n \cF_{\m\n} = 0 \; ,
\lab{A-eqs-0}
\ee
with
\be
a_2(M) \equiv 2F^{\ast 0} L^\pr (F^2)\bv_{F^2 = F^2(M)} = \mathrm{const}
\lab{a2-const} \; ;
\ee
\br
-\sqrt{\g^{(p)}} \pa_0 \(\chi \pa_0 X^\m\) +
\pa_i\(\chi\sqrt{\g^{(p)}} \g^{ij} \pa_j X^\m\)
\nonu \\
+ \chi\sqrt{\g^{(p)}} \(-\pa_0 X^\n \pa_0 X^\l + \g^{kl} \pa_k X^\n \pa_l X^\l\)
\G^\m_{\n\l} - q\, p! F^{\ast 0} \sqrt{\g^{(p)}} \pa_0 X^\n \cF_{\l\n}G^{\l\m}
\nonu
\er

\vspace{-0.6cm}

\br
- \frac{\b}{(p+1)!} \vareps^{a_1\ldots a_{p+1}} \pa_{a_1} X^{\m_1} \ldots
\pa_{a_{p+1}} X^{\m_{p+1}} \cF_{\l \m_1 \dots \m_{p+1}} G^{\l\m} = 0 \; .
\lab{X-eqs-0}
\er

\section{Lightlike Brane Dynamics in Static Gravitational Backgrounds}

Let us split the bulk space-time coordinates as:
\br
\( X^\m\) = \( x^a, y^\a\) \equiv \( x^0, x^i, y^\a\)
\lab{coord-split} \\
a=0,1,\ldots, p \;\; ,\;\; i=1,\ldots, p \;\; ,\;\;
\a = 1,\ldots, D-(p+1)
\nonu
\er
and consider static ($x^0$-independent) background metrics $G_{\m\n}$ of the
form:
\be
ds^2 = - A(y)(dx^0)^2 + C(y) g_{ij}(\vec{x}) dx^i dx^j + B_{\a\b}(y) dy^\a dy^\b
\; .
\lab{static-metric}
\ee

Here we will discuss the simplest non-trivial ansatz for the \textsl{LL-brane}
embedding coordinates:
\be
X^a \equiv x^a = \s^a \quad, \quad X^{p+\a} \equiv y^\a = y^\a (\t) \quad,
\quad \t \equiv \s^0  \; .
\lab{X-ansatz}
\ee
With \rf{static-metric} and \rf{X-ansatz}, the constraint Eqs.\rf{gamma-eqs-0} yield:
\be
- A(y(\t)) + B_{\a\b}(y(\t)) \ydot^\a \ydot^\b = 0 \quad ,\quad
C(y(\t)) g_{ij} - 2 a_1 (M) \g_{ij} = 0 \; ,
\lab{gamma-eqs-1}
\ee
where $\ydot^\a \equiv \frac{d}{d\t} y^\a$. Second Eq.\rf{gamma-eqs-1}
together with the last relation in \rf{gamma-p-const} implies:
\be
\frac{d}{d\t} C(y(\t)) = \ydot^\a\!\!\partder{}{y^\a}C \bv_{y=y(\t)}=0 \; .
\lab{C-rel}
\ee

The second-order Eqs.\rf{X-eqs-0} in the absence of couplings to bulk Maxwell
and Kalb-Ramond fields (which will be case we will consider in the present
section) yield accordingly:
\br
\pa_\t \chi + \frac{\chi}{A(y)} \ydot^\b\!\!\partder{}{y^\b} A(y) \bv_{y=y(\t)} = 0
\; ,
\lab{X0-eq} \\
\yddot^{\,\a} + \ydot^\b \ydot^\g \G^\a_{\b\g} + 
B^{\a\b} \(\frac{p\, a_1(M)}{C(y)}\partder{}{y^\b} C(y) + 
\h\partder{}{y^\b} A(y)\)\bv_{y=y(\t)}
\nonu \\
- \frac{\ydot^\a}{A(y)} \ydot^\b\!\!\partder{}{y^\b} A(y) \bv_{y=y(\t)} = 0 \; .
\lab{y-eqs}
\er
where $\G^\a_{\b\g}$ is the Christoffel connection for the metric $B_{\a\b}$
in the extra dimensions (cf. \rf{static-metric}). 

Here we will be interested in the case of constant brane tension:
\be
\pa_\t \chi = 0 \quad \to\quad 
\ydot^\a\!\!\partder{}{y^\a}A \bv_{y=y(\t)} = 0 \;\; 
\mathrm{from ~Eq.\rf{X0-eq}}\; .
\lab{A-rel}
\ee
Thus we arrive at the following compatible system of equations describing a
nontrivial motion of the \textsl{LL-brane} in the extra dimensions:
\br
\ydot^\a\!\!\partder{}{y^\a}A \bv_{y=y(\t)} = 0 \quad ,\quad
\ydot^\a\!\!\partder{}{y^\a}C \bv_{y=y(\t)}=0 \; ,
\lab{A-C-rel} \\
- A(y(\t)) + B_{\a\b}(y(\t)) \ydot^\a \ydot^\b = 0 \; ,
\lab{gamma-eqs-2} \\
\yddot^{\,\a} + \ydot^\b \ydot^\g \G^\a_{\b\g} + 
B^{\a\b} \(\frac{p\, a_1(M)}{C(y)}\partder{}{y^\b} C(y) + 
\h\partder{}{y^\b} A(y)\)\bv_{y=y(\t)} = 0
\lab{y-eqs-1}
\er

\subsection{Example 1: Two Flat Extra Dimensions}

In this case:
\br
y^\a = (\rho,\phi) \quad ,\quad 
B_{\a\b}(y) dy^\a dy^\b = d\rho^2 + \rho^2 d\phi^2 \; ;
\lab{flat} \\
A= A(\rho) \;\; ,\;\; C= C(\rho) \quad ;\quad \stackrel{.}{\rho} = 0 \;\;
,\;\; \mathrm{i.e.}\;\; \rho = \rho_0 = \mathrm{const} \; .
\lab{A-C-flat}
\er
Eqs.\rf{gamma-eqs-2} and \rf{y-eqs-1} yield correspondingly:
\br
- A(\rho_0) + \rho_0^2 \stackrel{.}{\phi}^2 = 0 \; ;
\lab{gamma-eqs-flat} \\
- \rho_0 \stackrel{.}{\phi}^2 + 
\(\frac{p\, a_1(M)}{C(\rho)} \pa_\rho C + \h \pa_\rho A)\)\bv_{\rho=\rho_0} = 0 
\quad ,\quad \stackrel{..}{\phi} = 0  \; .
\lab{y-eqs-flat}
\er
The last Eq.\rf{y-eqs-flat} implies:
\be
\phi (\t) = \om \t \; ,
\lab{rotation-flat}
\ee
which upon substituting into \rf{gamma-eqs-flat}--\rf{y-eqs-flat} gives:
\be
\om^2 = \frac{A(\rho_0)}{\rho_0^2} \quad ,\quad
A(\rho_0) = \rho_0 
\(\frac{p\, a_1(M)}{C(\rho)} \pa_\rho C + \h \pa_\rho A)\)\bv_{\rho=\rho_0} \; .
\lab{orbit-flat}
\ee
Thus, we find that the \textsl{LL-brane}
performs a planar circular motion in the flat extra
dimensions whose radius $\rho_0$ and angular velocity $\om$ are determined
from \rf{orbit-flat}. This property of the \textsl{LL-branes}
has to be contrasted with the usual case of Nambu-Goto-type braneworlds which
(in the ground state) occupy a fixed position in the extra dimensions.

\subsection{Example 2: Spherical Extra Dimensions}

In this case:
\br
y^\a = (\th,\phi) \quad ,\quad 
B_{\a\b}(y) dy^\a dy^\b = d\th^2 + \sin^2(\th) d\phi^2 \; ;
\lab{spherical} \\
A= A(\th) \;\; ,\;\; C= C(\th) \quad ;\quad \stackrel{.}{\th} = 0 \;\;
,\;\; \mathrm{i.e.}\;\; \th = \th_0 = \mathrm{const} \; .
\lab{A-C-spherical}
\er
Eqs.\rf{gamma-eqs-2} and \rf{y-eqs-1} yield correspondingly:
\br
- A(\th_0) + \sin^2 (\th_0) \stackrel{.}{\phi}^2 = 0 \; ;
\lab{gamma-eqs-spherical} \\
- \sin (\th_0) \cos (\th_0) \stackrel{.}{\phi}^2 + 
\(\frac{p\, a_1(M)}{C(\th)} \pa_\th C + \h\pa_\th A\)\bv_{\th=\th_0} = 0 \quad ,\quad
\stackrel{..}{\phi} = 0  \; .
\lab{y-eqs-spherical}
\er
Therefore, once again we obtain:
\be
\phi (\t) = \om \t \; ,
\lab{rotation-spherical}
\ee
which upon substituting into \rf{gamma-eqs-spherical}--\rf{y-eqs-spherical} gives:
\be
\om^2 = \frac{A(\th_0)}{\sin^2 (\th_0)} \quad ,\quad
A(\th_0) = \tan (\th_0)
\(\frac{p\, a_1(M)}{C(\th)} \pa_\th C + \h\pa_\th A\)\bv_{\th=\th_0} \; .
\lab{orbit-spherical}
\ee
As in the case of flat extra dimensions, Eqs.\rf{orbit-spherical} determine
the position $\th_0$ of the circular orbit of the \textsl{LL-brane}
and its angular velocity $\om$.

\subsection{Example 3: Toroidal Extra Dimensions}

In this case:
\be
y^\a = (\th,\phi)\;\; ,\;\; 0 \leq \th,\phi \leq 2\pi \quad ,\quad 
B_{\a\b}(y) dy^\a dy^\b = d\th^2 + a^2 d\phi^2 \; ;
\lab{toroidal}
\ee
Eqs.\rf{A-C-rel}--\rf{y-eqs-1} assume the form:
\br
\(\stackrel{.}{\th} \pa_\th A + 
\stackrel{.}{\phi} \pa_\phi A\)\bv_{\th=\th (\t),\,\phi=\phi (\t)} = 0 \; ,
\nonu \\
\(\stackrel{.}{\th} \pa_\th C + 
\stackrel{.}{\phi} \pa_\phi C\)\bv_{\th=\th (\t),\,\phi=\phi (\t)} = 0 \; ;
\lab{A-C-toroidal} \\
- A\bv_{\th=\th (\t),\,\phi=\phi (\t)} + \stackrel{.}{\th}^2 + 
a^2 \stackrel{.}{\phi}^2 = 0 \; ;
\lab{gamma-eqs-toroidal} \\
\stackrel{..}{\th} + 
\(\frac{p\, a_1(M)}{C} \pa_\th C + \h\pa_\th A\)\bv_{\th=\th (\t),\,\phi=\phi (\t)}
= 0 \; ,
\nonu \\
\stackrel{..}{\phi} + 
\(\frac{p\, a_1(M)}{C} \pa_\phi C + \h\pa_\phi A\)\bv_{\th=\th (\t),\,\phi=\phi (\t)}
= 0 \; .
\lab{y-eqs-toroidal}
\er
Eqs.\rf{A-C-toroidal} can be solved by taking $A(\th,\phi)$ and
$C(\th,\phi)$ as functions of only one combination $\xi(\th,\phi)$ such that:
\br
A = A\bigl(\xi(\th,\phi)\bigr) \quad ,\quad C = C\bigl(\xi(\th,\phi)\bigr)
\lab{A-C-toroidal-1} \\
\frac{d}{d\xi}A\bv_{\th=\th (\t),\,\phi=\phi (\t)} = 0 \quad ,\quad
\frac{d}{d\xi}C\bv_{\th=\th (\t),\,\phi=\phi (\t)} = 0 \; .
\lab{A-C-toroidal-2}
\er
Taking into account \rf{A-C-toroidal-2}, Eqs.\rf{y-eqs-toroidal} imply:
\be
\stackrel{..}{\th}=0 \;\; ,\;\; \stackrel{..}{\phi} = 0 \quad ,\quad
\mathrm{i.e.}\;\; \th (\t) = \om_1 \t \;\; ,\;\; \phi (\t) = \om_2 \t \; .
\lab{y-eqs-toroidal-1}
\ee
Furthermore, taking into account the periodicity of $A$ and $C$ w.r.t.
$(\th,\phi)$ we find:
\be
\xi(\th,\phi) = \th - N\phi \quad ,\quad \om_1 = N \om_2 \; ,
\lab{xi-eq}
\ee
where $N$ is abritrary positive integer. In other words, from 
\rf{A-C-toroidal-1}--\rf{A-C-toroidal-2}
the admissable form of the background metric must be of the form:
\be
A = A (\th - N\phi)\;\; ,\;\; C = C (\th - N\phi) \quad ,\quad
A^\pr (0) = 0 \;\; ,\;\; C^\pr (0) = 0 \; ,
\lab{A-C-toroidal-3}
\ee
whereas Eq.\rf{gamma-eqs-toroidal} determines the angular frequencies $\om_{1,2}$
in \rf{y-eqs-toroidal-1}:
\be
\om_1^2 = \frac{A(0)}{1 + a^2/N^2} \quad , \quad \om_2 = \frac{\om_1}{N} \; .
\lab{omega-eq}
\ee
A particular choice for $A$ (and similarly for $C$) respecting conditions
\rf{A-C-toroidal-2} is:
\be
A = A_0 \sin^2 \bigl(\frac{\th - N\phi}{2}\bigr) + A_1 \quad, \quad
A_{0,1} = \mathrm{positive ~const} \; .
\lab{A-choice}
\ee

Thus, we conclude that the \textsl{LL-brane} performs a spiral motion in the toroidal
extra dimensions with winding frequences as in \rf{omega-eq}.

\section{Outlook}

In the present paper we presented a systematic Lagrangian formulation of
lightlike $p$-branes in arbitrary $(p+1)$ world-volume dimensions allowing
in addition for natural (gauge-invariant) couplings to bulk electromagnetic
and Kalb-Ramond gauge fields. In the context of ``brane-world scenarios''
lightlike braneworlds (of codimension two or more) in their ground state 
perform non-trivial motions in the extra dimensions unlike ordinary
Nambu-Goto braneworlds which position themselves at certain fixed points in
the extra dimensions. 

The special case of codimension one \textsl{LL-branes} needs separate study
which is relegated to a subsequent paper. As already discussed in 
refs.\ct{will-brane-all} for lightlike membranes ($p\!=\!2$) in $D\!=\!4$ 
bulk space-time, the \textsl{LL-brane} dynamics dictates that
the bulk space-time must possess an event horizon which is automatically 
occupied by the \textsl{LL-brane} (an explicit dynamical realization of the 
``membrane paradigm'' in black hole physics \ct{membrane-paradigm}).
Extending our treatment in refs.\ct{will-brane-all},
we will study the important issue of self-consistent
solutions for bulk gravity-matter systems (\textsl{e.g.}, Einstein-Maxwell-type)
coupled to lightlike branes where the latter serves as a source for gravity,
electromagnetism, dynamically produces space-varying cosmological constant and
triggers non-trivial matching of two different space-time geometries  
across common event horizon spanned by the lightlike brane itself.

Let us mention the observation in ref.\ct{beciu}, that large extra
dimensions could be rendered undetectable (due to the zero eigenvalue of the
induced metric) if our world is considered as a lightlike brane moving in 
$D>4$ bulk space  --  precisely the brane-world scenario obtained in the 
present paper from the consistent unified dynamical (Lagrangian) description 
of lightlike branes

To stress again, in our formalism we consider the {\em intrinsic} metric 
$\g_{ab}$ on the
world-volume of the {\em lightlike} brane to be the metric that defines the
geometry experienced by an observer confined to the brane. This is in
contrast to the induced metric \rf{metric-short-hand}, which as a result of
lightlike nature of the brane is necessarily {\em singular}, having  
spacelike components and a zero eigenvalue, \textsl{i.e.} a lightlike instead of
timelike one. Nevertheless, it is possible to 
ascribe a physical role to singular induced metrics provided they possess an
additional timelike (diagonal) component. The latter can be achieved by considering
\textsl{LL-brane} with $(p+2)$-dimensional world-volume (cf. \rf{LL-brane})
embedded in a $D$-dimensional ($D\!>\!p+2$) bulk space with {\em two} timelike 
dimensions ($G_{\m\n}$ having signature $(-,-,+,\ldots,+)$). Repeating the steps in 
Section 2 we will get an induced $(p+2)\times (p+2)$ metric
\rf{metric-short-hand} with signature $(0,-,+,\ldots,+)$, \textsl{i.e.},
with one lightlike, one timelike and $p$ spacelike dimensions. Then,
applying the formalism for degenerate metrics proposed in ref.\ct{searight},
one can employ the resulting induced metric as a starting point for 
construction of a Kaluza-Klein model with the pertinent lightlike brane
(with $(p+2)$-dimensional world-volume) as
a total Kaluza-Klein space with naturally unobservable extra dimension
(the first lightlike one) from the point of view of the ``normal''
$(p+1)$ world-volume dimensions. Also, let us note that the use of an embedding 
space-time with two timelike coordinates has an advantage if we want 
to obtain a Lorentz-invariant ground state, since there is the
possibility of having one additional time, not involved in the motion of the
lightlike brane.

%


\section*{Acknowledgments}
E.N. is sincerely grateful to the organizers of the 7-th International
Workshop ``Lie Theory and Its Applications in Physics'', Varna (2007).
E.N. and S.P. are supported by European RTN network
{\em ``Forces-Universe''} (contract No.\textsl{MRTN-CT-2004-005104}).
They also received partial support from Bulgarian NSF grant \textsl{F-1412/04}.
Finally, all of us acknowledge support of our collaboration through the exchange
agreement between the Ben-Gurion University of the Negev (Beer-Sheva, Israel) and
the Bulgarian Academy of Sciences.


\end{document}